\documentstyle[12pt, fleqn]{article}
\def\ref{\par\noindent\hangindent=6mm\hangafter=1}

\begin{document}
\baselineskip=4.5mm
\

\vspace{10mm}

\begin{center}
\begin{bf}
ANISOTROPIES OF COSMIC BACKGROUND RADIATION\\
FROM A LOCAL COLLAPSE\\
\end{bf}

\vspace{30mm}

Xiang-Ping Wu\\
DAEC, Observatoire de Paris-Meudon\\
92195 Meudon Principal Cedex, France\\
and\\
Beijing Astronomical Observatory\\
Chinese Academy of Sciences, Beijing 100080, China\\
(e-mail: wxp@frmeu51.bitnet)\\

\medskip
and\\
\medskip

Li-Zhi Fang  \\
Steward Observatory and Department of Physics\\
University of Arizona, Tucson, AZ 85721\\
(e-mail: fanglz@time.physics.arizona.edu)\\

\bigskip
\bigskip
\bigskip

\hspace{40mm}

To appear in Astrophysical Journal April 1, 1994
\end{center}

\newpage

\

\begin{center}
{\bf Abstract}
\end{center}

We present an exact solution of the anisotropies of cosmic background
radiation (CBR) from a local collapse described by a spherical
over-dense region embedded in a flat universe, with the emphasis on
the relationship between the dipole $(\Delta {\sf T}/{\sf T})_d$ and
the quadrupole $(\Delta {\sf T}/{\sf T})_q$ anisotropy. This result
has been used to examine the kinematic quadrupole correction
$(\Delta {\sf T}/{\sf T})_q=(\Delta {\sf T}/{\sf T})_d^2/2$,
which is usually applied to remove the contamination of the quadrupole
produced by local density inhomogeneities when finding the cosmic
amplitude of the quadrupole at the surface of last scattering. We
have found that the quadrupole of local collapse origin cannot always
be approximately described by the kinematic quadrupole.
Our numerical result shows that the difference between
the kinematic and local quadrupoles depends on the size and matter density
in the peculiar field, and the position of the observer. For a given
dipole, the local quadrupole can be different from the kinematic
quadrupole by a factor as large as 3. Therefore, the kinematic quadrupole
correction remains an uncertain factor in the determination of the
amplitude of a cosmic quadrupole.  Nevertheless, a preliminary analysis
shows that this uncertainty might not dramatically change the cosmological
origin of the COBE-DMR's quadrupole, unless a huge peculiar gravitational
field is assumed.

\bigskip

{\noindent}{\it Subject headings:} cosmic background radiation - cosmology:
theory

\newpage

\vspace{5mm}

\begin{center}
{\bf 1. INTRODUCTION}
\end{center}

This paper is aimed at studying the contribution of local
collapse of matter to the anisotropies of cosmic background radiation
(CBR), especially
the relationship between the dipole and the kinematic quadrupole.

It is generally believed that the CBR dipole
anisotropy comes from a Doppler effect. If the observer moves with
velocity $v$ with respect to the CBR rest frame, the special
relativity (SR) Doppler effect
would lead to a frequency-independent thermodynamic temperature
distribution on CBR (Peebles \& Wilkinson, 1968)
\begin{equation}
{\sf T}={\sf T_0}\left[1+\frac{v}{c}\cos\Psi+\frac{1}{2}(\frac{v}{c})^2
                \cos2\Psi+\cdot\cdot\cdot\right]
\end{equation}
The terms in the right-hand side are,
respectively, the monopole, the dipole and the kinematic quadrupole
anisotropies.

Eq.(1) is often  used 1) to determine the peculiar velocity $v$, and 2)
to calculate the correction of the kinematic quadrupole. For instance,
the recent COBE-DMR first year sky maps show a dipole amplitude
$(\Delta{\sf T})_d= 3.365\pm0.027$ mK toward direction
$(l^{II}, b^{II} ) = (264.4^o\pm 0.3^o, 48.4^o\pm 0.5^o)$
(Smoot et al, 1992; Bennett et al, 1992; Kogut et al, 1993).
If the entirely observed dipole results from our peculiar
motion, the kinematic quadrupole anisotropy should be
$(\Delta{\sf T})_q=(\Delta{\sf T})^2_d/2{\sf T} \approx 2.1 \mu$K.
This result has been used by the COBE-DMR team
to obtain the CBR quadrupole anisotropy (Smoot et al, 1991, 1992).
They removed the kinematic quadrupole term from the DMR maps
in order to eliminate the influence of local density inhomogeneities on
the quadrupole amplitude. Since the amplitude of the kinematic
quadrupole is about 13\% of the cosmic quadrupole, the difference
between the quadrupoles produced by the SR kinematics and local density
inhomogeneities would be one of the factors leading to uncertainty in the
amplitude of the cosmological quadrupole. Therefore, it is necessary
to study the conditions, under
which the kinematic quadrupole correction given by eq.(1) is valid.

The initial velocity of cosmic matter underwent a decrease due to
the expansion of the universe. It can then be reasonably
assumed that our present peculiar velocity results totally from the infall
motion toward the center of the local collapse of matter, i.e.,
the peculiar motion of the observer is completely given by the
gravitation of
local matter clustering. In linear approximation,
the relationship between the matter density fluctuation
$\delta$ and the peculiar velocity is (Peebles, 1980)
\begin{equation}
v=\frac{1}{3}\left(H_0D\right)\Omega^{0.6}\delta
\end{equation}
where $D$ is the distance  to the center of the
perturbation, $\Omega$, the density parameter and $H_0$, the Hubble
constant. Thus, eqs.(1) and (2) imply that the CBR dipole and
kinematic quadrupole anisotropies are essentially caused by the
local density inhomogeneities. In other word, the CBR's dipole
and quadruple anisotropy should be calculated as an effect of
a locally time-dependent density inhomogeneity.

The question is then: can the contribution of a local
collapse of matter to the CBR dipole and quadrupole anisotropies be
approximately described by a Doppler effect of eq.(1), i.e. a pure
SR effect? Obviously, the
SR Doppler effect description is principally different
from that of a local collapse. The former is a kinematic effect,
of which the only parameter is the observer's velocity $v$. The
latter is a dynamical model, and governed by the matter distribution or the
initial density  contrast $\delta$, the
location of the observer and the size of the local collapse.

Eqs.(1) and (2) also tell us that the dipole
anisotropy depends on the first order of the density fluctuation $\delta$
and then the kinematic quadrupole is of the second order.
It is well known from general relativity that in a linear approximation
the behavior of a comoving object in an expansion or collapsing metric
can be equivalently described as a Doppler motion. But such an equivalence
will no longer be held if the higher orders are involved. Indeed,
in terms of the dipole term one can not distinguish between a SR
Doppler effect and
a local collapse. The velocity $v$ determined by a
dipole anisotropy is just equal
to the $v$ caused by the locally gravitational collapse.
However, one can not expect that the
SR kinematic quadrupole remains the same as the quadrupole caused by
the local collapse.

The original intention of the kinematic quadrupole correction in the COBE
sky maps is to remove the contamination of a quadrupole given by the
local density inhomogeneity. Therefore, in principle, this correction
should not be done by the kinematic quadrupole, but the {\it local}
collapse of matter. We need then to conduct a quantitative comparison of
the kinematic quadrupole with that of a local collapse.

The CBR anisotropies from a collapse at redshift greater than 1
have been addressed by several authors (e.g. Rees \& Sciama, 1968;
Dyer, 1976; Olson \& Silk; 1979; Raine \& Thomas, 1981;
Kaiser, 1982; Occhionero, Santangelo \& Vittorio,
 1983; Nottale, 1984; Dyer \& Ip, 1988; Arnau et al, 1993).
However, the contribution from a local collapse to the CBR
anisotropies, especially the relationship between the dipole and local
quadrupole, has not been carefully studied thus far. We will
investigate  the CBR dipole and quadrupole
anisotropies from a locally spherical collapse described by a
Tolman-Bondi universe. This model has recently been used
to analyze the CBR anisotropy produced by perturbations with sizes
of $100\sim1000$ Mpc (Fang and Wu, 1993, hereafter Paper I).
The advantage of this model is that one can
find all needed exact solutions, which allow us to
check the availability of the kinematic quadrupole correction of eq.(1).

In Section 2, we will discuss the metric and null geodesic in a spherically
symmetric collapse. Section 3 shows the linear, the high-order
and the exact numerical solutions of the dipole and quadrupole anisotropies.
In Section 4, a relationship between peculiar velocity and CBR
anisotropies will be presented. Finally, a brief summary and
discussion will be given in Section 5.

\bigskip

\begin{center}
{\bf 2. METRIC AND NULL GEODESIC}
\end{center}

For simplicity, we consider a spherical density perturbation embedded
in a flat universe. The universe can then be generally modeled
as a Tolman-Bondi metric (Paper I):
\begin{equation}
ds^2= e^{\lambda(x,t)}dx^2+r^2(x,t)(d\theta^2+\sin^2\theta d\phi^2)-dt^2
\end{equation}
and
\begin{equation}
e^{\lambda(x,t)}=\frac{r'^2}{1+x^2H^2_i(1-\frac{\overline{\rho}(x,t_i)}
{\rho_{ci}})}
\end{equation}
where  $H_i$, $\overline{\rho}(x,t_i)$
and $\rho_{ci}$ are, respectively, the Hubble constant, the density and the
critical density of the universe at epoch $t_i$.
We denote that $'=\partial /\partial x$ and
$\dot{ }=\partial/\partial t$.

The dynamics of the collapse from the initial perturbation at $x=0$
is described by the collapse factor $S(x,t)$ which is defined as
\begin{equation}
r=S(x,t)x,  \hspace{3mm} S(x,t_i)=1
\end{equation}
The dynamical equation of $S(x,t)$ is
\begin{equation}
\dot{S}^2-H_i^2\frac{\overline{\rho}(x,t_i)}{\rho_{ci}}\frac{1}{S}
=H_i^2(1-\frac{\overline{\rho}(x,t_i)}{\rho_{ci}})
\end{equation}
Let the initial density perturbation be $\delta(x)$, the density
distribution $\overline{\rho}(x,t_i)$ in eqs.(4) and (6) can then be
written as
\begin{equation}
\overline{\rho}(x,t_i)=\rho_{ci}(1+\delta(x))
\end{equation}
If the initial density perturbation $\delta_0$ is assumed to be
constant in the range of $x<x_c$, one has
\begin{equation}
\begin{array}{l}
\delta(x)=\left\{\begin{array}{ll}
\delta_0,  & x\leq x_c\\
\delta_0(x_c/x)^3, & x>x_c.\\
\end{array}\right.
\end{array}
\end{equation}
In this case, the dynamical equation (6) can be solved analytically
(Paper I).

The 0th-component of the null geodesic in the metric of eq.(3) is
\begin{equation}
\frac{dk^0}{d\sigma}=-\frac{1}{2}e^{\lambda}\dot{\lambda}
\left(\frac{dx}{d\sigma}\right)^2 - r \dot{r} \left(
\frac{d\phi}{d\sigma}\right)^2
\end{equation}
where $k^0$ is the 0th-component of a photon's four momentum and
$\sigma$ is an affine parameter of the null geodesic. From the
condition of $ds=0$, one can find the energy shift
of a photon, which is assumed to be emitted at $t=t_e$ with
frequency $\nu_e$ and received at $(x_0,t_0)$ with frequency $\nu_0$,
\begin{equation}
\frac{\nu_e}{\nu_0}=\exp\left(
\int_{t_e}^{t_0}\frac{1}{2}\dot{\lambda}dt\;+\;
\int_{t_e}^{t_0}(\frac{\dot{r}}{r}-\frac{1}{2}\dot{\lambda})
r^2\left(\frac{d\phi}{dt}\right)^2dt\right)
\end{equation}
The trajectories of the photon can be obtained from
\begin{equation}
\frac{dx}{dt}=\pm e^{-\lambda/2}\sqrt{1-r^2\left(\frac{d\phi}{dt}\right)^2}
\end{equation}
\begin{equation}
\frac{d^2\phi}{dt^2}+\left[\frac{2}{r}\left(\frac{dr}{dt}\right)
-\frac{1}{2}\dot{\lambda}\right]\frac{d\phi}{dt}
=\left(\frac{\dot{r}}{r}-\frac{1}{2}\dot{\lambda}\right)
r^2\left(\frac{d\phi}{dt}\right)^3
\end{equation}
Considering the trajectory of a photon is approaching a straight line when
$r$ is large, one can obtain  $d\phi/dt$ from solving the eq.(12), and then
the energy shift eq.(10) and the trajectory eq.(11) become
\begin{equation}
\frac{\nu_e}{\nu_0}=e^{\int_1^{T_0}\frac{1}{2}\dot{\lambda}t_edT}
\left(1-\int_1^{T_0}U t_edT\right)^{1/2}
\end{equation}
\begin{equation}
\frac{dX}{dT}=\pm e^{-\lambda/2} \left(1-
\frac{\xi}{1-\int_T^{T_0}Ut_edT}\right)^{1/2}
\end{equation}
where the new time and space coordinates ($T$, $X$)  are defined by
$T=t/t_e$ and $X=x/t_e$. Therefore, $T_0=t_0/t_e$, $X_0=x_0/t_e$ and
$X_c=x_c/t_e$.  $U$ and $\xi$ in eqs.(13) and (14) are, respectively,
\begin{equation}
U  = 2\xi(\frac{\dot{\lambda}}{2}-\frac{\dot{r}}{r})
\end{equation}
\begin{equation}
\xi  = \left(\frac{r_0}{r}\right)^2 \sin^2\Psi e^{\int_{T_0}^{T}
\dot{\lambda}t_e dT}
\end{equation}
with $r_0=r(x_0,t_0)$, and  $\Psi$ is the incidence angle of the photon,
i.e. the angle between the directions of the photon's trajectory and
the line of sight from observer to the center of the perturbation,

Since the universe is flat, the CBR anisotropy produced by the local
collapse is given by
\begin{equation}
\frac{\Delta{\sf T}}{\sf T}=\frac{\nu_0}{\nu_e}T_0^{2/3}-1
\end{equation}
where $T_0=(1+z_d)^{3/2}$ and $z_d$ is the redshift at decoupling
time ($t_e$). Without a loss of generality, we will
take $t_i=t_e$, i.e., the $\delta(x)$ is the density fluctuation
at the recombination time $t_e$.

Because we are interested in the effect of a local collapse,
in the following, only the case of $x_0 < x_c$ or $X_0 <X_c$ will
be considered.
Using the expansion of $S(X,T)$ given in the Appendix, one find
the solution of $\Delta {\sf T}/{\sf T}$ up to the second order
of $\delta_0$ as
\begin{eqnarray*}
\frac{\Delta{\sf T}}{\sf T}= -\int_1^{T_0}f_1\delta dT
                    +X_0^2 \sin^2\Psi\int_1^{T_{c0}}
                    \frac{u_1}{{X^{(0)}}^2}\delta dT
\end{eqnarray*}
\begin{eqnarray*}
\ \ \ +\frac{1}{2}\left[\int_1^{T_0}f_1\delta dT -
    X_0^2\sin^2\Psi\int_1^{T_{c0}}\frac{u_1}
    {{X^{(0)}}^2}\delta dT\right]^2
    +3\int_1^{T_{c0}}f_1\frac{\Delta X}{X^{(0)}}\delta dT
\end{eqnarray*}
\begin{eqnarray*}
\ \ \  -5 X_0^2 \sin^2\Psi\int_1^{T_{c0}}\frac{u_1}{{X^{(0)}}^2}
   \frac{\Delta X}{X^{(0)}}\delta dT
   +\left[3f_1(T_{c0})+
    X_0^2\sin^2\Psi\frac{u_1(T_{c0})}{X_c^2}\right]\delta_0\Delta T_{c}
\end{eqnarray*}
\begin{eqnarray*}
\ \ \   -\int_1^{T_0}f_2\delta^2 dT+
   X_0^2\sin^2\Psi\int_1^{T_{c0}}\frac{u_2}{{X^{(0)}}^2}\delta^2 dT
\end{eqnarray*}
\begin{eqnarray*}
\ \ \ +2\delta_0X_0^2\sin^2\Psi\frac{S_1(T_0)}{T_0^{2/3}}
    \int_1^{T_{c0}}\frac{u_1}{{X^{(0)}}^2}\delta dT
   -2X_0^2\sin^2\Psi\int_1^{T_{c0}}\frac{S_1(T)}{T^{2/3}}
    \frac{u_1}{{X^{(0)}}^2}\delta^2 dT
\end{eqnarray*}
\begin{equation}
\ \ \  +2X_0^2\sin^2\Psi\int_1^{T_{c0}}\frac{u_1}{{X^{(0)}}^2}\delta
    \int_{T_0}^{T}f_1\delta dT
    + X_0^4\sin^4\Psi\left[\int_1^{T_{c0}}
\frac{u_1}{{X^{(0)}}^2}\delta dT\right]^2
\end{equation}
where $f_i$ and $u_i$ ($i=1, 2$) are defined in the Appendix.
$\Delta X$ is the first-order correction to the photon trajectory,
and $\Delta T_c$, the cross time correction when the photon enters
into the perturbation regime. They are, respectively,
\begin{equation}
\Delta X  =  \pm \frac{\sqrt{(X^{(0)})^2-X_0^2\sin^2\Psi}}{X^{(0)}}
               \int_T^{T_0}\frac{F(X^{(0)},T)}{T^{2/3}}dT
\end{equation}
\begin{equation}
\Delta T_c  =  T_{c0}^{2/3}\int_{T_{c0}}^{T_0}\frac{F(X^{(0)},T)}{T^{2/3}}dT
\end{equation}
where $X^{(0)}$ is the zero-order solution of the photon's trajectory,
 which can
be found from  eq.(14) by taking $\delta=0$. It is
\begin{equation}
(X^{(0)})^2=X_0^2 \sin^2\Psi+(3 T_0^{1/3}-3T^{1/3}-X_0 \cos\Psi)^2
\end{equation}
Similarly, the zero-order solution to the photon cross time ($T_c$) is
found to be
\begin{equation}
T_{c0}=\left[T_0^{1/3}-\frac{X_0 \cos\Psi}{3}-\frac{1}{3}\sqrt{X_c^2-
X_0^2\sin^2\Psi}\right]^3.
\end{equation}
The function $F(X,T)$ in eqs.(19) and (20) is defined as
\begin{eqnarray*}
F(X,T)= g_1T^{2/3}\delta -\frac{X_0^2\sin^2\Psi}{X^2-X_0^2\sin^2\Psi}
\end{eqnarray*}
\begin{equation}
\ \ \ \ \ \ \ \;\;\;\;   \left[\frac{S_1(T_0)}{T_0^{2/3}}\delta_0
  -\frac{S_1(T)}{T^{2/3}}\delta
  +\int_{T_0}^T f_1\delta dT
  +X_0^2\sin^2\Psi\int_T^{T_0}\frac{u_1}{X^2}\delta dT
  \right]
\end{equation}
and $g_1$ is also given in the Appendix.

\bigskip

\begin{center}
{\bf 3. SOLUTIONS OF $\Delta {\sf T}/{\sf T}$}
\end{center}

\begin{center}
{\it 3.1 First-Order Solution}
\end{center}

The first-order solution of $\Delta{\sf T}/{\sf T}$ can be obtained by
substituting  the zero-order solution of the photon
trajectory $X^{(0)}$ of eq.(21) into the
first two terms of eq.(18). After a straightforward computation,
the first-order solution is found to be
\begin{equation}
\frac{\Delta{\sf T}}{\sf T}=\delta_0\left(\frac{X_c^2}{15}-
        \frac{X_0^2}{45}+\frac{2}{15}T_0^{1/3}X_0\cos\Psi
        -\frac{2}{135}\frac{X_c^3}{T_0^{1/3}}+O(T_0^{-2/3})\right)
\end{equation}
The above expression is also an expansion with respect to the
parameter $(1/T_0)$. The largest term is of the order of
$(1/T_0)^{-1/3}$. Eq.(24) shows that in the approximation up to
the first-order of $\delta_0$, $\Delta{\sf T}/{\sf T}$ consists mainly of
two parts: a monopole term
\begin{equation}
\left(\frac{\Delta{\sf T}}{\sf T}\right)_0 \simeq
\left(\frac{X_c^2}{15}-\frac{X_0^2}{45}-\frac{2}{135}\frac{X_c^3}{T_0^{1/3}}
\right)\delta_0
\end{equation}
and a dipole term
\begin{equation}
\left(\frac{\Delta{\sf T}}{\sf T}\right)_d\simeq
\frac{2}{15}T_0^{1/3}X_0\delta_0 \cos\Psi
\end{equation}

The physical explanations of these results are simple.
The monopole is given by the gravitational
redshift of the local matter inhomogeneity $\delta_0$, which leads
to an isotropic increase of the CBR temperature. The dipole
term depends on the distance between the observer and the
center of the density perturbation ($X_0$). In the case of $X_0=0$,
i.e., the observer sits at the center of the local collapse,
the dipole term will disappear. This indicates that the dipole anisotropy
is indeed due to the asymmetry of the local collapse around the observer.
However, it should be pointed out that the dipole anisotropy depends on
the time-dependence of the local inhomogeneity. If the
gravitation field around
the observer is static, the asymmetry ($X_0 \neq 0$) do not produce
such a dipole anisotropy.

Comparing eq.(26) with eq.(1), we have
\begin{equation}
v=\frac{2}{15} T_0^{1/3}X_0\delta_0
\end{equation}
This means that the CBR dipole anisotropy produced by a local collapse
can be equivalently described as a SR Doppler effect if the observer
is assumed to have a peculiar velocity given by eq.(27).
Let's show that eq.(27) is indeed the observer's peculiar velocity produced
by the gravitation field of the local collapse. The proper distance
$D$ corresponding to $X=x/t_e$ is
\begin{equation}
D=\frac{2}{3}\frac{c}{H_0}\frac{1}{\sqrt{1+z_i}}X
\end{equation}
where $z_i$ is the redshift when the perturbation occurred. We will take
it to be the redshift at decoupling era, i.e. $z_i = z_d$.
In the paper I, we have found that for a given $\delta_0$ the present
density contrast of the local collapse should be
\begin{equation}
\frac{\Delta\rho}{\rho} \approx \frac{3}{5}(1+z_d)\delta_0.
\end{equation}
In a flat universe, $\rho = \rho_{cr} = 3H_0^2/8\pi G$.
Using eqs.(28) and (29),  eq.(27) can be rewritten as
\begin{equation}
v= \frac{1}{3}\left(H_0D\right)\left(\frac{\Delta \rho}{\rho}\right)
=\frac{2}{3H_0}g
\end{equation}
where $g=G\Delta M/D^2$ is the observer's acceleration raised
by the extra-mass $\Delta M=(4\pi/3)D^3\Delta\rho_0$. Therefore,
$v$ in eq.(27) is the same as that in eq.(2). In a word,
in the linear approximation it is reasonable to determine the observer's
peculiar velocity by the CBR dipole anisotropy and the Doppler effect
formula (1).

If the CBR dipole is totally given by the local collapse, one can then
find from eq.(26) that
$X_0\delta_0 = (15/2)(\Delta {\sf T}/{\sf T})_dT_0^{-1/3} \sim 2.6\times
10^{-4}$. This is, in fact,  a constrain to the local
collapse causing the dipole. For instance, if we assume the size of this local
collapse is of the order of the distance to the Great Aractor, i.e.
$X_0 \sim 1$, the initial density perturbation $\delta_0$ in this area
should be about $\delta_0 \sim 2 \times 10^{-4}$, or from eq.(29)
today's density contrast is about $1 \times 10^{-1}$.

\smallskip

\begin{center}
{\it 3.2 Second-Order Solutions and Quadrupole Anisotropy}
\end{center}

In a similar way, the second-order solution of
$\Delta {\sf T}/{\sf T}$ can be found from eq.(18) as follows
\begin{flushleft}
\begin{equation}
\frac{\Delta{\sf T}}{\sf T}= \delta_0^2\left[(\frac{3}{175}X_c^2
-\frac{11}{1575}X_0^2)T_0^{2/3}+\frac{4}{175}T_0X_0\cos\Psi
+\frac{2}{225}T_0^{2/3}X_0^2\cos2\Psi\right]
\end{equation}
\end{flushleft}
which is also written in the series of $(1/T_0)$ up to the order
of $(1/T_0)^{-2/3}$. Eq.(31) indicates that, up to the $\delta_0^2$
approximation, the dominant term is still the dipole, because the
amplitude of dipole is of the order of $T_0$, while the amplitudes
of the monopole and quadrupole terms are only of the order of
$T_0^{2/3}$.

The quadrupole anisotropy of the local collapse now is
\begin{equation}
\left(\frac{\Delta{\sf T}}{\sf T}\right)_q=
\frac{2}{225}T_0^{2/3}X_0^2\delta_0^2 \cos2\Psi
\end{equation}
Comparing this amplitude with that in eq.(26), one find
\begin{equation}
\left(\frac{\Delta{\sf T}}{\sf T}\right)_q
=\frac{1}{2}\left(\frac{\Delta{\sf T}}{\sf T}\right)_d^2
\end{equation}
This is just the SR relationship between the anisotropies of the
dipole and kinematic quadrupole. Therefore, the SR kinematic quadrupole
correction is available till the approximation of eq.(32) is correct.
However, when the terms of the order of $T_0^{1/3}$, $T_0^{0}$ are
taken into account, the dipole-quadrupole relation of eq(33)
will no longer hold true. For instance, up to the order of
$\delta_0^{2}$ and $T_0^{1/3}$, eq.(33) should be corrected as
\begin{equation}
\left(\frac{\Delta{\sf T}}{\sf T_0}\right)_q =
\frac{1}{2}\left(\frac{\Delta{\sf T}}{\sf T}\right)_d^2
+ T_0^{1/3}X_0^2\Delta_q\delta_0^2
\end{equation}
where the factor $\Delta_q$ is given by
\begin{eqnarray*}
\Delta_q= -\frac{19X_c}{3780}-{\frac{1}{X_0}}\left[\frac{X_0}{140X_c}
+\frac{229X_0^3}{61440X_c^3}+\frac{261X_0^5}{81920X_c^5}
+\frac{3X_0^7}{4096X_c^7}\right]
\end{eqnarray*}
\begin{equation}
\ \ \ \ \ \ +X_0 \left[\frac{41X_0}{9800X_c}
 -\frac{1333X_0^3}{2064384X_c^3}
+\frac{467X_0^5}{5734400X_c^5}+\frac{3833X_0^7}{11468800X_c^7}
+O(\frac{X_0^9}{X_c^9})\right]
\end{equation}
Eq.(35) shows that the term of the SR kinematic quadrupole does not
always dominate the quadrupole anisotropy shown in eq.(34).
Therefore, the SR kinematic quadrupole may not always be a good
approximation of the quadrupole produced by a local collapse.
In principle, the kinematic quadrupole correction of
eq.(33) should also be replaced by the local quadrupole correction of
eq.(34).

Let's define a ratio between the local quadrupole to
the SR kinematic quadrupole as
\begin{equation}
q=\frac{(\Delta {\sf T}/{\sf T})_q}
{(1/2)(\Delta {\sf T}/{\sf T})^2_d}
\end{equation}
Figure 1 plots the dependence of $q$ on $X_c$ and $X_0$ as given by eq.(35).
One can see that $q$ does not sensitively depend on the observer
distance but the whole size of the local inhomogeneity. The
difference between local and kinematic quadrupole, i.e.
that $q$ significantly deviates from 1, mainly
occurs in two cases: 1) very small clustering with a scale less than
 a few Mpc, and 2) very large clustering with a scale greater than
$10^3$ Mpc.

\smallskip

\begin{center}
{\it 3.3 Numerical Solution}
\end{center}

In order to accurately compare the kinematic quadrupole with the local
collapse model, we made the numerical solutions of the dipole and
quadrupole amplitude of the local collapse model.
These solutions depend on three parameters: 1) the
initial density fluctuation $\delta_0$, 2) the size of
local collapse $X_c$, and 3) the distance of the observer to the
center of the local collapse $X_0$.

We still lack the detailed knowledge of
the local collapse. It is generally believed that the bulk motion
of horizon sized volume is negligible. Therefore, the Local Group's
peculiar velocity, which may govern the dipole term, was probably induced
by the local inhomogeneities in the density field with size comparable
with horizon at the time of decoupling.
We have shown in Paper I that the COBE-DMR
result of CBR anisotropy on an angular scale of $10^{\circ}$
implies the existence of collapses on scale as large as
about 1000 h$_{50}^{-1}$ Mpc with an
initial density fluctuation $\delta_0 \sim (2.8 - 6.9)\times10^{-6}$,
 or its present density enhancement is $(1.7 - 4.2)\times 10^{-3}$.
On the other hand, the distance of our Local Group to the center
of the collapse should at least be greater than the distance to the
Great Attractor, which is estimated to be 80 h$_{50}^{-1}$Mpc
(Lynden-Bell et al, 1988), i.e. $X_0=0.7$.
Therefore, it would be valuable to consider the following two cases:
$X_c = 1.4$ ($\sim$ 150 h$_{50}^{-1}$Mpc) and 10
($\sim$ 1000 h$_{50}^{-1}$Mpc),
i.e., the lower value of $X_c$ is about 2 times of the distance to the
Great Attractor and the higher value of $X_c$ is about the size of horizon.

The numerical results are listed in Table 1, in which [{\it exact}]
indicates the exact
solution of the dipole amplitude, and $[\delta_0]$ and
$[\delta_0^2]$ denote, respectively, the solutions up to the first- and
second-order of $\delta_0$. The density fluctuations at the decoupling
epoch ($z_d=10^{3}$) are taken to be $10^{-3}$,
$10^{-4}$ and $10^{-5}$, respectively. The corresponding values of
the SR kinematic quadrupole
$(1/2)(\Delta{\sf T}/{\sf T})_d^2= (1/2)[exact]^2$ have also been
listed for comparisons. The term of $(\Delta{\sf T}/{\sf T})_q$ is
the exact solution of the quadrupole amplitude.

One can find from Table 1 that: a) For the dipole anisotropy the
linear approximation is good  if the initial density
fluctuation is less than $10^{-4}$. b) The dipole is nearly independent
of the size of the local collapse. c) The local quadrupole sensitively
depends on the size of the perturbation. d) The kinematic quadrupole
is always larger than the exact solution of the local quadrupole. e)
 The difference
between the kinematic and exact quadrupoles does not vanish with
the decrease of density perturbation $\delta_0$. In the case of $X_c=10$
the ratio $1/q$ can be as large as 3.

\bigskip

\begin{center}
{\bf 4. DIPOLE AS THE FUNCTIONS OF DENSITY CONTRAST AND PECULIAR VELOCITY}
\end{center}

In this section, we intend  to represent the CBR's dipole anisotropy by the
observable parameters such as the present density contrast and
peculiar velocity. Up to the second order of $\delta_0$ the dipole
anisotropy is (see eqs.(24)
and (31))
\begin{equation}
\left(\frac{\Delta{\sf T}}{\sf T}\right)_d=
\left(\frac{2}{15}T_0^{1/3}\delta_0+\frac{4}{175}T_0\delta_0^2 + ...
\right)X_0\cos\Psi
\end{equation}

First, the present density contrast $\Delta\rho/\rho$
as a function of initial fluctuation $\delta_0$ is given by
\begin{equation}
\frac{\Delta\rho}{\rho}=\left(\frac{S_0}{S(X,T_0)}\right)^3
(1+\delta_0)-1
\end{equation}
Using eq.(A1--A5), one can find the expression of $\Delta \rho/\rho$
expanded as a series of $\delta_0$
\begin{eqnarray*}
\frac{\Delta\rho}{\rho}= \delta_0\left[\frac{3}{5}T_0^{2/3}
+\frac{2}{5}T_0^{-1}\right]
\end{eqnarray*}
\begin{eqnarray*}
\ \ \ \ \ \  +\delta_0^2\left[\frac{51}{175}T_0^{4/3}-\frac{2}{5}T_0^{2/3}
      +\frac{4}{25}T_0^{-1/3} -\frac{6}{35}T_0^{-1}
      +\frac{3}{25}T_0^{-2}\right]
\end{eqnarray*}
\begin{eqnarray*}
\ \ \ \ \ \  +\delta_0^3\left[\frac{341}{2625}T_0^2
    -\frac{68}{175}T_0^{4/3}+\frac{1}{3}T_0^{2/3}+\frac{34}{875}T_0^{1/3}
     -\frac{92}{525}T_0^{-1/3}
     +\frac{2}{21}T_0^{-1}\right.
\end{eqnarray*}
\begin{equation}
\ \ \ \ \ \  \left. + \frac{14}{375}T_0^{-4/3}
    -\frac{18}{175}T_0^{-2}
     +\frac{4}{125}T_0^{-3}\right]
     + O(\delta_0^4)
\end{equation}
Reversing this expansion and retaining the first three terms, we have
\begin{equation}
\delta_0 \approx \frac{1}{1+z_d}\left(\frac{\Delta\rho}{\rho}\right)
\left[\frac{5}{3}-\frac{85}{63}\left(\frac{\Delta\rho}{\rho}\right) +
\frac{14075}{11907}\left(\frac{\Delta\rho}{\rho}\right)^2 +...
\right]
\end{equation}
Substituting this result into eq.(37), one find the expression of dipole
up to the second-order of the present density contract $\Delta\rho/\rho$,
\begin{equation}
\left(\frac{\Delta{\sf T}}{\sf T}\right)_d=
\frac{1}{3}\left(\frac{H_0D}{c}\right)\left(\frac{\Delta\rho}{\rho}\right)
\left(1-\frac{11}{21}\frac{\Delta\rho}{\rho}\right).
\end{equation}
Therefore, in linear approximation, the dipole anisotropy would be
overestimated from the measurement of the local density fluctuation.

Second, the peculiar velocity inside the perturbation regime can be
derived from the continuity equation (Peebles, 1980)
\begin{equation}
\frac{\partial\Delta\rho/\rho}{\partial T}+
\frac{1}{S_0}\nabla\cdot (1+\Delta\rho/\rho){\bf v}=0.
\end{equation}
It is
\begin{equation}
v=\frac{1}{2}\left(H_0D\right)T_0^{1/3}
\left(\frac{\partial\Delta\rho/\rho}{\partial T}\right)
\frac{S_0}{1+\Delta\rho/\rho}
\end{equation}
Using eqs.(38), (39) and (40), the peculiar velocity can be expanded
by $\Delta\rho/\rho$ as
\begin{equation}
v\approx\left(H_0D\right)
        \left[\frac{1}{3}\left(\frac{\Delta\rho}{\rho}\right)-
        \frac{4}{63}\left(\frac{\Delta\rho}{\rho}\right)^2+
        \frac{328}{11907}\left(\frac{\Delta\rho}{\rho}\right)^3+
        \cdot\cdot\cdot\right]
\end{equation}
This result is consistent with the recent work by Gramann (1993), who
showed that the peculiar velocity would be overestimated by the linear
approximation.
Substituting this expression into eq.(37), we have the solution
of the dipole anisotropy as a function of peculiar velocity
\begin{equation}
\left(\frac{\Delta{\sf T}}{\sf T}\right)_d=\frac{c}{v}
\left[1+\frac{25}{63}\left(\frac{c}{DH_0}\right)\left(\frac{v}{c}\right)
+...\right]
\end{equation}
or
\begin{equation}
\frac{v}{c}=\left(\frac{\Delta{\sf T}}{\sf T}\right)_d/\left(1+
\frac{25}{198}\frac{\Delta\rho}{\rho}\right)
\end{equation}
Therefore, the SR relation eq.(1)
can be used to determine peculiar velocity from the CBR measurement
only if the local density contrast $\Delta\rho/\rho$ is very small.
Considering the fact that the present density contrast $\Delta\rho/\rho$
is of the order of 1 on the scale of superclusters, the higher
order correction for the peculiar field may not be negligible.

The peculiar velocity divergence in our model is simply
\begin{equation}
\nabla\cdot{\bf v}=-\dot{S_0}\left(\frac{\Delta\rho}{\rho}\right)
\left[1-\frac{4}{21}\left(\frac{\Delta\rho}{\rho}\right)
+\frac{328}{3969}\left(\frac{\Delta\rho}{\rho}\right)^2+...\right]
\end{equation}
or
\begin{equation}
\nabla\cdot{\bf v}= - \frac{\dot{S_0}(\Delta\rho/\rho)}
{1+0.190(\Delta\rho/\rho)-0.046(\Delta\rho/\rho)^2+...}
\end{equation}
A similar result has been empirically found recently by Nusser and Dekel
(1991, 1993).

\bigskip

\begin{center}
{\bf 5. DISCUSSION AND CONCLUSIONS}
\end{center}

It is usually believed that in linear approximation the CBR
anisotropies consist primarily of three parts, namely, 1)
a Doppler effect from the motion of the
observer with respect to the CBR rest frame, 2) a Sachs-Wolfe effect
at the surface of last scattering and 3) a time-dependent potential
effect along the photon path (Mart\'inez-Gonz\'alez, Sanz \& Silk,
1990). In this description, one can not, in fact, distinguish
between the effects of a time-dependent potential and a Doppler
motion.  The contribution of a local collapse of matter to CBR
anisotropy can totally be contained in the Doppler effect of the
observer's peculiar velocity.
One can then use the linear relationship between the peculiar velocity $v$
and local density contrast $\Delta\rho/\rho$ to calculate the CBR dipole
anisotropy $(\Delta{\sf T}/{\sf T})_d$ [eq.(1)].

However, in terms of the CBR quadrupole anisotropy,
the equivalence between the Doppler effect and the
local gravitational field  will no longer exists.
This is because
kinematic quadrupole is essentially non-linear.
As a result, the quadrupole anisotropy produced by a local inhomogeneity
is not generally equal to the SR kinematic quadrupole.
Especially, if the peculiar field is comparable with the
horizon size, the relationship between the
dipole and the quadrupole terms will be substantially different from that
given by the SR Doppler effect.
We have found that, based on a simply spherical density perturbation model,
the local quadrupole can be different from the kinematic quadrupole by
a factor as large as 3.

The purpose of the SR kinematic quadrupole
correction used in the data reduction of COBE observation is to remove
the local quadrupole amplitude from the
CBR sky temperature maps, and then the remaining maps will contain only the
component of the cosmic quadrupole, i.e. the component given by
the density fluctuation on the last scattering surface. The
kinematic quadrupole correction is, in fact, the correction of local
quadrupole. Yet, as we have shown, the local  quadrupole
can not be simply replaced by the SR kinematic quadrupole. Therefore, the
original purpose of the kinematic
quadrupole correction may  not be completely achieved by removing the
quadrupole given by eq.(1).

The quadrupole amplitude of the local density inhomogeneities depends
on the matter distribution,
the density contrast, the size of the local gravitational field and the
position of the observer. All these parameters are poorly
known at the moment. Therefore, it seems to be very hard to precisely identify
the local quadrupole. A possible way to obtain the local
 quadrupole
would be to analyze the temperature map to see if they contain a component
with polar axis paralleled to the direction of dipole. This is because
the axes of both
the dipole and local quadrupole should point out to the center of the
local inhomogeneity. Principally, without
a precise local quadrupole correction, the observed
quadrupole should be regarded as a sum of cosmological and local
components. Thus, the local quadrupole correction leads to an
uncertain factor in finding the cosmological quadrupole term
from observed CBR maps (e.g., the COBE measurement).

In spite of the fact that the local quadrupole may be remarkably
different from the kinematic quadrupole, our simple model shows that
for a given dipole amplitude the amplitude of the local quadrupole
is always less than the corresponding kinematic quadrupole,
if  the local peculiar field is of the order of $100$ Mpc in size.
This means that
the kinematic quadrupole seems to be an upper limit to the
local quadrupole. Because the presently
observed
quadrupole amplitude by COBE is about 8 times larger than this
upper limit, one may concludes that  the uncertainty of the local
quadrupole would not lead
to an uncertainty greater than 13\% of the observed cosmic quadrupole.

Certainly, this conclusion is model-dependent. The present model
is actually a toy one, which has assumed the
simplest matter distribution -- a constant density contrast. It
may largely deviate from the real local  matter distribution.
For instance, the peculiar velocity distribution derived from
this toy model [eq.(44)] shows a linear increase with the
distance from the perturbation center, which does not  fit
with the observations around the Great Attractor (Faber \&
Burstein, 1989). Moreover, if the initial density perturbation
is assumed to be highly non-spherical, the local quadrupole may
have an amplitude greater than $(\Delta{\sf T}/{\sf T})^2_d$.
Therefore, the loophole hidden in the kinematic quadrupole correction
has not been totally closed yet. One needs to simulate this effect
by using a more realistic model of local collapse of matter, e.g.,
the models given by N-body simulations. The study of the
interaction between the local matter distribution and the CBR quadrupole
would be of great significance for a better understanding of
both the initial perturbation and the local collapse of matter.

\bigskip
\bigskip

The authors thanks Dr. J. Heine for his helps. WXP is grateful to the
CNRS and the Wong K.C. Foundation for financial support.  This work
was also partially supported by NSF contract INT 9301805.

\bigskip

\setcounter{equation}{0}

\renewcommand{\theequation}{A\arabic{equation}}

\begin{center}
{APPENDIX}
\end{center}

For the initial perturbation eq.(8), the collapse factor $S(x,t)$ can be
expanded in the series of $\delta_0$ as
\begin{equation}
S(X,T)=S_0(T)+S_1(T)\delta_0+S_2(T)\delta_0^2+S_3(T)\delta_0^3 + ...
\end{equation}
where the coefficients $S_i$ ($i=0,1,2,\cdot\cdot\cdot$) can be determined
by eq.(6) to be
\begin{equation}
S_0(T)  = T^{2/3}
\end{equation}
\begin{equation}
S_1(T)  = -\frac{2}{15 T^{1/3}}+\frac{T^{2/3}}{3}
-\frac{T^{4/3}}{5}
\end{equation}
\begin{equation}
S_2(T)  = -\frac{1}{225 T^{4/3}}+\frac{4}{315 T^{1/3}}
        +\frac{4 T^{1/3}}{75} -\frac{T^{2/3}}{9}
        +\frac{T^{4/3}}{15} - \frac{3 T^2}{175}
\end{equation}
\begin{eqnarray*}
S_3(T)  = -\frac{4}{10125 T^{7/3}}+\frac{11}{4725 T^{4/3}}
        -\frac{2}{1125 T^{2/3}} + \frac{2}{945 T^{1/3}}
        -\frac{64 T^{1/3}}{1575}
\end{eqnarray*}
\begin{equation}
\ \ \ \ \ \ \ \ \ \ \ +\frac{5 T^{2/3}}{81}+\frac{6 T}{875}
         -\frac{2 T^{4/3}}{45}
          +\frac{3 T^2}{175} -  \frac{23 T^{8/3}}{7875}
\end{equation}
Similarly, one can find the expansions of $r$,  $\lambda$,
$\xi$ and $U$ with respect to $\delta_0$ by their definitions (Paper I)
\begin{equation}
\dot{\lambda}=\frac{2\dot{r}'}{r'}
\end{equation}
and
\begin{equation}
\frac{1}{2}\dot{\lambda}=\frac{x\dot{S}'+\dot{S}}{xS'+S}
\end{equation}
as well as eqs.(15) and (16). Let
\begin{equation}
\frac{\dot{\lambda}t_e}{2}  =  \frac{2}{3T} +f_1\delta_0 +f_2\delta_0^2+ ...
\end{equation}
\begin{equation}
\frac{\dot{\lambda}}{2}-\frac{\dot{r}}{r}  =  u_1\delta_0 +u_2\delta_0^2+ ...
\end{equation}
\begin{equation}
e^{-\lambda/2}  =  \frac{1}{T^{2/3}}+ g_1\delta_0 +g_2\delta_0^2+...
\end{equation}
where the coefficients $f_i=f_i(T)$, $u_i=u_i(T)$ and
$g_i=g_i(X,T)$ ($i=1,2,3,\cdot\cdot\cdot$) are found to be

{\noindent}1. for $X<X_c$,
\begin{equation}
f_1  =  \frac{2}{15 T^2}-\frac{2}{15 T^{1/3}}
\end{equation}
\begin{equation}
f_2  =  \frac{2}{75T^3}-\frac{2}{35T^2}-\frac{2}{225T^{4/3}}
          +\frac{4}{45 T^{1/3}}-\frac{26 T^{1/3}}{525}
\end{equation}

\smallskip

\begin{equation}
u_i  =  0, \;\;\;\;\; i=1,2,3,\cdot\cdot\cdot
\end{equation}

\smallskip

\begin{equation}
g_1  =  \frac{1}{5}+\frac{2}{15 T^{5/3}}-\frac{1}{3 T^{2/3}}
          -\frac{2X^2}{9 T^{2/3}}
\end{equation}
\begin{eqnarray*}
g_2  =  -\frac{1}{5}+\frac{1}{45 T^{8/3}}-\frac{32}{315 T^{5/3}}
          +\frac{2}{9 T^{2/3}}+\frac{2 T^{2/3}}{35}
\end{eqnarray*}
\begin{equation}
\ \  \ \ \ \  -\frac{2X^2}{45}
          -\frac{4X^2}{135T^{5/3}}+\frac{2X^2}{27T^{2/3}}
          -\frac{2X^4}{81T^{2/3}}
\end{equation}

\smallskip

{\noindent}2. for $X>X_c$,

\begin{equation}
f_1  =  -\frac{4}{15 T^2}+\frac{4}{15 T^{1/3}}
\end{equation}
\begin{equation}
f_2  =  \frac{2}{75T^3}-\frac{4}{35T^2}+\frac{28}{225T^{4/3}}
          -\frac{2}{45 T^{1/3}}+\frac{4 T^{1/3}}{525}
\end{equation}

\smallskip

\begin{equation}
u_1  =  -\frac{2}{5T^2}+\frac{2}{5T^{1/3}}
\end{equation}
\begin{equation}
u_2  =  -\frac{2}{35T^2}+\frac{2}{15T^{4/3}}-\frac{2}{15T^{1/3}}
          +\frac{2T^{1/3}}{35}
\end{equation}

\smallskip

\begin{equation}
g_1  =  -\frac{2}{5}-\frac{4}{15 T^{5/3}}+\frac{2}{3 T^{2/3}}
          -\frac{2X^2}{9 T^{2/3}}
\end{equation}
\begin{eqnarray*}
g_2  =  -\frac{1}{5}+\frac{11}{225 T^{8/3}}-\frac{92}{315 T^{5/3}}
          +\frac{12}{25T}-\frac{1}{9 T^{2/3}}
\end{eqnarray*}
\begin{equation}
\ \ \ \ \ \   +\frac{13 T^{2/3}}{175}
          +\frac{4X^2}{45}
          +\frac{8X^2}{135T^{5/3}}-\frac{4X^2}{27T^{2/3}}
          -\frac{2X^4}{81T^{2/3}}
\end{equation}

\newpage

\begin{small}
\begin{flushleft}

\begin{tabular}{c|c|c|c|c|c|c}
\multicolumn{7}{c}{Table 1 Comparison of the SR Kinematic Quadrupole with
the Exact Solutions} \\
\multicolumn{7}{c}{\  \ }\\
\hline
 size  & $\delta_0$ & \multicolumn{3}{|c|}{$(\Delta {\sf T}/{\sf T})_d$}
 & $[(\Delta {\sf T}/{\sf T})_d]^2/2$ & $(\Delta {\sf T}/{\sf T})_q$\\
\cline{3-5}
 &  & $[\delta_0]/\delta_0$ & $([\delta_0]+[\delta^2])/\delta_0$ &
[exact]/$\delta_0$ & $\times \delta_0^{-2}$ & $\times \delta_0^{-2}$ \\
\hline
  &  &  &  &  &  &  \\
           & $10^{-3}$ & 2.951 & 3.457 & 3.638 & 6.6 &  6.0 \\
 $X_c=1.4$ & $10^{-4}$ & 2.951 & 3.002 & 3.003 & 4.5 &  4.1 \\
           & $10^{-5}$ & 2.951 & 2.957 & 2.957 & 4.4 &  4.1 \\
  &  &  &  &  &  &  \\
\hline
  &  &  &  &  &  &  \\
           & $10^{-3}$ & 2.951 & 3.457 & 3.632 & 6.6 &  2.6 \\
 $X_c=10 $ & $10^{-4}$ & 2.951 & 3.002 & 3.000 & 4.5 &  1.4 \\
           & $10^{-5}$ & 2.951 & 2.957 & 2.953 & 4.4 &  1.4 \\
  &  &  &  &  &  &  \\
\hline
\end{tabular}
\end{flushleft}
\end{small}

\newpage

\begin{center}
{\bf REFERENCES}
\end{center}

\ref Arnau, J.V., Fullana, M.J., Monreal, L., \& Saez, D. 1993,
           ApJ, 402, 359

\ref Bennett, C.L. et al 1992, ApJ, 396, L7

\ref Dyer, C.C. 1976, MNRAS, 175, 429

\ref Dyer, C.C. \& Ip, P.S.S. 1988, MNRAS, 235, 895

\ref Faber, S.M. \& Burstein, D. 1989, in {\it Large-Scale Motion
           in the Universe}, eds. V.C.Rubin \& G.V. Coyne,
           (Princeton:  Princeton University  Press)

\ref Fang, L.Z. \& Wu, X.P. 1993, ApJ, 408, 25 (Paper I)

\ref Gramann,M. 1993, ApJ, 405, L47

\ref Kaiser, N. 1982, MNRAS, 198, 1033

\ref Kogut, A. et al 1993, ApJ, 1993, submitted

\ref Lynden-Bell,D., Faber,S.M., Burstein,D., Davies,R.L.,
           Dressler,A., Terlevich,R.J. \& Wegner,G. 1988,
           ApJ, 326, 19

\ref Mart\'inez-Gonz\'alez, E., Sanz, J.L. \& Silk, J. 1990, ApJ, 355, L5

\ref Nottale,L. 1984, MNRAS, 206, 713

\ref Nusser,A. \& Dekel,A. 1991, ApJ, 379, 6

\ref Nusser,A. \& Dekel,A. 1993, ApJ, 405, 437

\ref Occhionero, F. Santangelo, P. \& Vittorio, N. 1983,
A\&A, 177, 365

\ref Olson, D.W. \& Silk, J. 1979, ApJ, 233, 395

\ref Peebles, P.J.E. 1980, in {\it The Large-Scale Structure of
           the Universe}, ed. A.S.Wightman \& P.W.Anderson (Princeton:
           Princeton University Press)

\ref Peebles, P.J.E. \& Wilkinson, D.T. 1968, Phys.Rev., 174, 2168

\ref Raine, D.J. \& Thomas, E.G. 1981, MNRAS, 195, 649

\ref Rees, M.J. \& Sciama, D.W. 1968,  Nature, 217, 511

\ref Smoot, G.F. et al 1991, ApJ, 371, L1

\ref Smoot, G.F. et al 1992, ApJ, 396, L1

\newpage

\begin{center}
Figure Captions
\end{center}

{\noindent}{\bf Figure 1.} The ratio $q$ of the local quadrupole
$(\Delta {\sf T}/{\sf T})_q$ to the SR kinematic quadrupole
$(1/2)({\Delta {\sf T}}/{\sf T})_d^2$.
Only the terms with orders greater than $T_0^{1/3}$ in eqs.(34)
are considered. This plot illustrates the dependence of
the ratio $q$ on the size of the inhomogeneity and the position of
the observer. For the collapse with intermediate scale of about
$10^2$ Mpc, we have $q\approx1$. For both the very small and very large
scale collapses, the local quadrupole will significantly be different
from the kinematic quadrupole.

\end{document}